\documentclass[useAMS,usenatbib,usegraphicx]{mn2e}
\usepackage{color}

\title[Alignment in star-debris disc systems seen by Herschel]
{Alignment in star-debris disc systems seen by Herschel}

\author[J. S. Greaves et al.]{J. S. Greaves$^1$
, G. M. Kennedy$^2$, N. Thureau$^1$, C. Eiroa$^3$, J. P. Marshall$^3$, 
\newauthor{J. Maldonado$^3$, B. C. Matthews$^{4,5}$, G. Olofsson$^6$, M.J. Barlow$^{7}$, 
A. Moro-Mart\'{i}n$^8$, }
\newauthor{B. Sibthorpe$^{9,10}$, O. Absil$^{11}$, D. R. Ardila$^{12}$, M. Booth$^5$, 
H. Broekhoven-Fiene$^5$, }
\newauthor{D.J.A. Brown$^1$, A. Collier Cameron$^1$, C. del Burgo$^{13}$, 
J. Di Francesco$^{4,5}$, G. D\^{u}chene$^{14}$, }
\newauthor{J. Eisl\"{o}ffel$^{15}$, S. Ertel$^{16}$, W.S. Holland$^9$, J. Horner$^{17}$, 
P. Kalas$^{14,18}$, J.J. Kavelaars$^{4,5}$, }
\newauthor{J.-F. Lestrade$^{19}$, L. Vican$^{20}$, D.J. Wilner$^{21}$, S. Wolf$^{22}$, 
M.C. Wyatt$^2$ } \\
$^1$SUPA, School of Physics and Astronomy, University of St. 
Andrews, North Haugh, St. Andrews KY16 9SS, UK \\
$^2$Institute of Astronomy, University of Cambridge, Madingley 
Road, Cambridge CB3 0HA, UK \\
$^3$Dpt. F\'{i}sica Te\'{o}rica, Facultad de Ciencias, Universidad Aut\'{o}noma 
de Madrid, Cantoblanco, 28049 Madrid, Spain \\
$^4$National Research Council of Canada, 5071 West Saanich Road., Victoria, BC, Canada, 
V9E 2E7, Canada \\
$^5$University of Victoria, Finnerty Road, Victoria, BC, V8W 3P6 Canada  \\
$^6$Stockholm Observatory, SCFAB, SE-106 91 Stockholm, Sweden \\
$^{7}$Department of Physics and Astronomy, University College London, Gower Street, 
London WC1E 6BT, UK \\
$^8$Centro de Astrobiología (CSIC-INTA), 28850 Torrej\'{o}n de Ardoz, Madrid, Spain \\
$^9$UK Astronomy Technology Center, Royal Observatory, Blackford Hill, Edinburgh 
EH9 3HJ, UK \\
$^{10}$SRON, Postbus 800, 9700 AV Groningen, The Netherlands \\
$^{11}$Dept. d'Astrophysique, G\'{e}ophysique et Oc\'{e}anographie, Universit\'{e} 
de Li\`{e}ge, 17 all\'{e}e de Six Ao\^{u}t, B-4000 Sart-Tilman, Belgium \\
$^{12}$NASA Herschel Science Center, IPAC, MS 100-22, California Institute of 
Technology, Pasadena CA 91125, USA \\
$^{13}$Instituto Nacional de Astrof\'{i}sica, \'{O}ptica y Electr\'{o}nica, Luis 
Enrique Erro 1, Sta. Ma. Tonantzintla, Puebla, Mexico \\
$^{14}$Astronomy Department, University of California, Berkeley, CA 94720 \\
$^{15}$Th\"{u}ringer Landessternwarte, Sternwarte 5, 07778, Tautenburg, Germany \\
$^{16}$IPAG, Universit\'{e} Joseph Fourier / CNRS, 414 Rue de la Piscine, 38400 
St-Martin d'H\`{e}res, France \\
$^{17}$Department of Astrophysics and Optics, School of Physics, University of New South 
Wales, Sydney, 2052 Australia \\
$^{18}$SETI Institute, Mountain View, CA 94043, USA \\
$^{19}$Observatoire de Paris - CNRS, 77 Av. Denfert Rochereau, 75014 Paris, France \\
$^{20}$Department of Physics and Astronomy, University of California, Los Angeles, 
CA 90095, USA \\
$^{21}$Harvard-Smithsonian Center for Astrophysics, 60 Garden Street, Cambridge, MA 02138, 
USA \\
$^{22}$Institut f\"{u}r Theoretische Physik und Astrophysik, Universit\"{a}t zu Kiel, 
Liebnizstr. 15, 24118 Kiel, Germany 
}

\begin{document}

\date{Accepted 2013. Received 2013; in original form 2013}

\pagerange{\pageref{firstpage}--\pageref{lastpage}} \pubyear{2011}

\maketitle

\label{firstpage}

\begin{abstract}

Many nearby main-sequence stars have been searched for debris using the far-infrared 
{\it Herschel} satellite, within the DEBRIS, DUNES and Guaranteed-Time Key Projects. 
We discuss here 11 stars of spectral types A to M where the stellar inclination is 
known and can be compared to that of the spatially-resolved dust belts. The discs are 
found to be well aligned with the stellar equators, as in the case of the Sun's 
Kuiper belt, and unlike many close-in planets seen in transit surveys. The ensemble 
of stars here can be fitted with a star-disc tilt of $\la 10^{\circ}$. These results 
suggest that proposed mechanisms for tilting the star or disc in fact operate rarely. 
A few systems also host imaged planets, whose orbits at tens of AU are 
aligned with the debris discs, contrary to what might be expected in models where 
external perturbers induce tilts.

\end{abstract}


\section{Introduction}

The planets in the Solar System orbit near a plane aligned with the Sun's equator. This 
is tilted by only $7^{\circ}$ with respect to the ecliptic plane (Beck \& Giles 2005), 
with the midplane of the more dynamically-excited Kuiper Belt aligned within $2^{\circ}$ 
of the ecliptic (Brown \& Pan 2004; Collander-Brown et al. 2003). However, many asteroids 
have very inclined orbits, attributed to scattering by planets or to the Kozai mechanism 
(dynamical exchange of high eccentricities and inclinations). Such effects are of renewed 
interest with the discovery of extremely inclined orbits of some extrasolar planets, 
including cases so extreme as to be retrograde (e.g. Brown et al. 2012, Simpson et al. 
2011, Triaud et al. 2010, Winn et al. 2010). These bodies are observed in transit, where 
the occulting planet blocks starlight with specific Doppler shifts (the 
Rossiter-McLaughlin effect). It is widely thought that perturbations from more distant 
(unseen) planets allow the Kozai mechanism to operate, or lead to mutual scattering, and 
potentially tidal orbital circularisation and stellar spin-axis reorientation (Winn et 
al. 2010; Albrecht et al. 2012).

Here we explore whether distant planetesimals can have orbits misaligned with the 
stellar spin axis. It has been proposed that interaction of the magnetic field of 
a young star with its circumstellar disc could tip the star (Foucart \& Lai 2011; 
Lai et al. 2011). Alternatively, external accretion could give a randomised final 
angular momentum vector to the disc (Bate et al. 2010), or encounters with another 
disc/envelope system could cause dynamical perturbation (Thies et al. 2011). 
Evidence of such events could be found much later, for main-sequence stars 
where belts of planetesimals have formed from the circumstellar discs, as  
collisions generate debris that produces thermal emission at infrared and longer 
wavelengths. Further, where planets have been imaged or detected by astrometry, the 
inclinations of the orbital and belt planes can be compared to the stellar equator. 

Results of star-disc alignment studies are so far sparse. Greaves et al. (2004) 
noted that the nearby old Solar-analogue $\tau$~Ceti appeared to have a rather edge-on 
debris disc while the star's small projected rotational velocity ($v \ sin \ i_*$) 
suggested a more pole-on aspect. However, confusion with background objects hinders 
inclination estimation for this compact disc (Di Francesco et al., in prep.). Watson et 
al. (2011) examined 8 debris systems with Sun-like host stars, but found no cases where 
the disc and star were definitely misaligned. However, the data available spanned a wide 
range of wavebands and angular resolutions, potentially causing problems where 
interferometers resolved out disc flux, or dispersed small grains 
were seen in scattered light. It is therefore timely to make an update 
using newly-resolved discs from surveys made with the large and sensitive {\it Herschel} 
observatory (Pilbratt et al. 2010). The PACS camera (Poglitsch et al. 2010) provided  
uniform imaging at 5.6-11.4 arcsecond resolution at wavelengths of 70, 100, 160 $\umu$m. 
We identify here 11 main-sequence stars (some planet-hosting) that now have resolved 
debris discs along with information on the stellar inclination. The relative alignments 
are then compared to theoretical expectations.

\section{Data analysis}

\subsection{Disc Data}

\begin{figure}
\label{fig1}
\includegraphics[height=42mm]{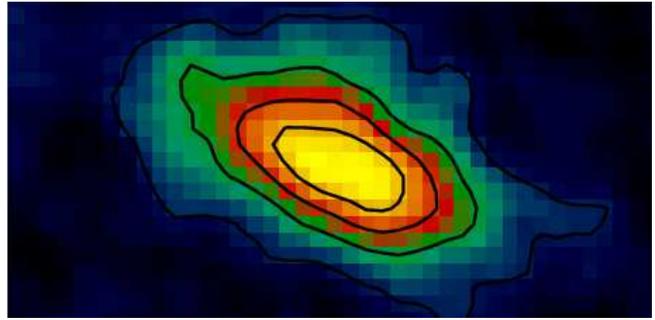}
\caption{DEBRIS image of the HD~115617 (61~Vir) system at 70~$\umu$m, in a 40$\times$20~arcsec 
RA, Dec. field with 5.6~arcsec beam, after subtracting stellar emission. For this 
8th-closest G-dwarf to the Sun, 10 arcseconds corresponds to 85~AU. Image from Wyatt et 
al. (2012; Figure~2); see this paper for model details. 
}
\end{figure}

{\it Herschel} data were obtained for Key Projects awarded under Guaranteed Time 
(Olofsson 2010) and for the larger unbiased Open Time Key Programmes DEBRIS (Matthews et 
al. 2010) and DUNES (Eiroa et al. 2010, 2013). For spectral types AFGKM, debris detection 
rates are up to $\sim$25\%, with numerous discs now spatially resolved with PACS (Booth 
et al. 2013; Eiroa et al. 2013). As an example, Figure~1 shows data for HD~115617 
(61~Vir), where the debris disc is nearly edge-on, and a less inclined disc would appear 
distinctly rounder. Implicitly, we assume that the discs have negligible vertical 
thickness and are circular; Greaves et al. (in prep.) discuss this in the context of 
highly resolved discs. HR~8799 (Herschel PI project; Matthews, in prep.) 
was added to the final sample; this interesting planet-host system was previously 
resolved by {\it Spitzer} at 70 $\umu$m (Su et al. 2009).

We fitted model discs to estimate inclinations with respect to the sky plane, so 
e.g. an $i_d = 0^{\circ}$ disc is face-on (while an $i_* = 0^{\circ}$ star is pole-on). 
The discs were analysed uniformly, with the two or three available wavebands fitted 
simultaneously, and least-squares minimisation was used to optimise the radius, position 
angle and inclination of a model thin annulus (see Wyatt et al. 2012 for description). 
Uncertainties in these $i_d$ values were estimated by comparing alternate inclinations 
obtained from a thin-toroid grid-search algorithm (Booth et al. 2013) and from 
beam-deconvolved 2-D Gaussian fits to the discs. Comparison of the outcomes shows an 
average 7$^{\circ}$ discrepancy between methods. Here we adopt a conservative error of 
$\pm10^{\circ}$ (at the upper end of measured differences), or the estimates from 
published detailed models. These include studies by Sibthorpe et al. (2010) for Vega, 
Wyatt et al. (2012) for 61~Vir, Broekhoven-Fiene et al. (2013) for $\gamma$~Dor, Lestrade 
et al. (2012) for GJ~581, while Marshall et al. (in prep.) will further discuss HD 30495 
and HD 110897. Models for Vega, 61~Vir and $\gamma$~Dor showed that discrepancies between 
fits at different wavelengths are small, with inclination estimates varying by only 
$\sim5^{\circ}$. As earlier spectral types tend to have better resolved discs 
(Booth et al. 2013; Eiroa et al. 2013), we subsequently order the systems by spectral type 
and then distance (Table~1; Figure~2), as a guide to the increasing difficulty of  
fitting inclinations.

\begin{table*}
 \caption{Debris disc systems observed by the {\it Herschel} Key Projects DEBRIS 
($\gamma$~Dor, HD 115617, GJ~581), DUNES (HD110897, HD 30495, HD 166, HD 17925, HD 
131511) and in Guaranteed Time (Vega, $\epsilon$~Eri, AU Mic), plus HR~8799 observed 
separately. Stars are listed by most common name, and ordered by spectral type and 
distance (noted by `UNS' identification where applicable; Phillips et al. 2010). 
Subsequent columns list system components, observed rotation periods $P$, projected 
rotation velocities $v \ sin \ i_*$ and radii $R_*$. Derived stellar inclinations are 
given by $i_*$, and fitted disc inclinations by $i_d$ (0$^{\circ}$ = pole- or face-on). 
The Sun (for comparison) has angles with respect to the ecliptic plane. 
Absolute value of inclination 
differences $|\Delta i|$ are $|i_*-i_d|$, with an uncertainty from errors in $i_d, i_*$ 
added in quadrature and adopting $(i_{*,min}+90^{\circ})/2$ in lower-limit cases; 
$sin^2 i_* + cos^2 i_d$ is a measure of misalignment (see text). Stellar periods are 
from: [1] Simon \& Fekel (1987); [2] Baliunas et al. (1996); [3] Gaidos et al. (2000); 
[4] Baliunas et al. (1983); [5] de Warf et al. (2010); [6] Donahue et al. (1996); 
[7] Bohigas et al. (1986); [8] Henry et al. (1995); [9] Vogt et al. (2010); [10] 
Messina et al. 2001; [11] Hebb et al. 2007.
}
 \label{tab:list}
 \begin{tabular}{@{}lllccccccc}
\hline
system names (UNS id)	& notes			& $P$		& $v \ sin \ i_*$& $R_*$& $i_*$ 	& $i_d$		& $|\Delta i|$	& $sin^2 i_*$		\\
			&			& (days)	& (km/s) & ($R_{\odot}$)& ($^{\circ}$)	& ($^{\circ}$)	& ($^{\circ}$) 	& $+ cos^2 i_d$		\\
\hline
Vega, HD 172167 (A003)	& planet?; 2 belts	& ---		& ---		& ---	& 3--6		& $10 \pm 2$    & $5.5 \pm 2.5$	& $0.98 \pm 0.01$ 	\\
HR 8799, HD 218396 (A---)& planets; 2 belts	& ---		& ---		& ---   & $\ga40$	& $27 \pm 10$	& $\ga 3$	& $\ga1.07$		\\
10 CVn, HD 110897 (F050)& 			& 13 [1] 	& $3.4 \pm 1.4$ & 0.99	& $63 (\geq 33)$& $56 \pm 10$ 	& $7_{(-7)}^{+29}$& $1.11 \pm 0.65$	\\
$\gamma$~Dor, HD 27290 (F085)& 2 belts		& ---		& ---		& ---	& 63--80	& $69 \pm 5$	& $3_{(-3)}^{+10}$	& $1.03 \pm 0.11$ 	\\
{\it Sun (G---)}	& {\it planets, 2 belts}& {\it ---}	& {\it---}	& {\it---}& ${\it 7.3}$	& ${\it 1.7 \pm 0.2}$	& ${\it 5.6 \pm 0.2}$ & ${\it 1.02 \pm 0.00}$	\\ 
61 Vir, HD 115617 (G008)& planets		& 29 [2]	& $1.6 \pm 0.5$	& 0.97	& $68 (\geq 41)$& $77 \pm 4$ 	& $9_{(-9)}^{+22}$& $0.91 \pm 0.52$	\\
58 Eri, HD 30495 (G029)	& 			& 11.3 [2,3,4]	& $3.4 \pm 0.3$ & 0.97	& $51 \pm 6$ 	& $51 \pm 10$ 	& $0_{(-0)}^{+12}$  	& $1.00 \pm 0.20$	\\
V439 And, HD 166 (G030)	& 2 belts? 		& 5.7 [3,5]	& $4.8 \pm 0.7$ & 0.87	& $39 \pm 6$ 	& $50 \pm 10$  	& $11_{(-11)}^{+12}$ 	& $0.81 \pm 0.20$	\\
$\epsilon$ Eri, HD 22049 (K001)	& planet(s); 2 belts& 11.6 [6]	& $2.3 \pm 0.3$ & 0.74	& $46 \pm 8$ 	& $38 \pm 10$ 	& $8_{(-8)}^{+13}$	& $1.14 \pm 0.22$	\\
EP Eri, HD 17925 (K035)	& 			& 6.9 [2,6,7]	& $5.8 \pm 0.6$ & 0.79	& $88 (\geq 63)$& $54 \pm 10$	& $34_{-27}^{+10}$& $1.34 \pm 0.25$	\\ 
DE Boo, HD 131511 (K053)& 			& 10.4 [8]	& $4.5 \pm 0.4$ & 0.91	& $\geq 70$ 	& $84 \pm 10$ 	& $4_{(-4)}^{+12}$& $1.06 \pm 0.18$	\\
HO Lib, GJ 581 (M056)	& planets		& 94 [9]	& $0.3 \pm 0.3$ & 0.30	& $\geq 0$    	& $50 \pm 20$	& ---		& ---			\\
AU Mic, HD 197481 (M---)& 			& 4.9 [10,11]	& $8.5 \pm 0.6$ & 0.77	& $\geq 81$	& $\geq 80$	& $1_{(-1)}^{+7}$	& $1.13 \pm 0.16$	\\
\hline
 \end{tabular}
\end{table*}

\subsection{Stellar Data}

Inclinations of stars are difficult to determine. In principle, interferometry of 
features of the stellar surface could give full 3-D information on the angle at which we 
view the star, the same as obtained from resolved disc images. However, even with 
ultra-high resolution this technique is mainly applicable to giant stars. Here only Vega 
has $i_*$ from interferometry; its apparent oblateness is sensitive to viewing angle 
because it is flattened by rapid rotation. Vega is very close to pole-on (Aufdenberg et 
al. 2006; Peterson et al. 2006; Yoon et al. 2010; Monnier et al. 2012), which minimises 
apparent oblateness, while our analysis of four other DEBRIS A/F-stars ($\beta$~Leo, 
$\alpha$~CrB, $\beta$~UMa, $\eta$~Crv) gave only weak lower limits to $i_*$. Stars seen 
nearly side-on are suggested when $v \ sin \ i_*$ approaches the maximum value for the 
spectral type, but this also has poor accuracy and is subject to the assumption that 
stars of a given spectral type have a maximum spin rate. This method was used only to check 
inclinations. Estimates of $i_*$ can also be made from models of asteroseismological data 
and/or rotation of spot patterns, as some surface features can only be seen in certain 
orientations. Here asteroseismology gives useful checks for HR~8799 (Wright et al. 2011), 
$\epsilon$~Eri (Croll et al. 2006, Fr\"{o}hlich 2007) and $\gamma$~Dor (Balona et al. 
1996).

The primary method remains the classic approach of Campbell \& Garrison (1985), 
yielding inclination of the stellar pole with respect to the line of sight when true 
rotation velocity can be compared to $v \ sin \ i_*$. This gives 
\begin{equation} 
sin \ i_* = 0.0198 \ P \ v \ sin \ i_* \ / \ R_*, 
\end{equation} 
\noindent where stellar rotation period $P$ is in days, projected rotation velocity 
$v \ sin \ i_*$ is in km/s and stellar radius $R_*$ is in solar radii. Radii are 
from fitting optical and near-infrared fluxes for luminosity and effective 
temperature, with interferometric measurements for $\epsilon$~Eri and GJ~581 (Di 
Folco et al. 2004; von Braun et al. 2011). Checks on radii using 
surface-brightness relations (Kervella et al. 2004) show differences only at the 
5~\% level. Thus for radius and also period (see below), uncertainties usually 
contribute negligibly to the error estimate in inclination, and Table~1 only lists the 
uncertainty in $i_*$ derived from that in $v \ sin \ i_*$. Then by differentiation, 
$\delta i_* = \delta(sin \ i_*) / cos \ i_*$, with $\delta(sin \ i_*) = 0.0198 \ P \ 
\delta(v \ sin \ i_*) \ / \ R_*$ from Eq.~1. In some cases, allowed values of $i_*$ 
range from a lower bound up to 90$^{\circ}$, and then the lower bound quoted is from 
$sin \ i_*$ minus its error.

Projected rotational velocities of stars are found by fitting their spectral lines, with 
modest differences between methods and calibration systems that have been well 
characterised by G\l\c{e}bocki \& Gnaci\'{n}ski (2005a). Here we compile values from 
G\l\c{e}bocki \& Gnaci\'{n}ski (2005b) plus $v \ sin \ i_*$ data from the subsequent 
literature, including a comprehensive study made for DUNES (Mart\'{i}nez-Arn\'{a}iz et 
al. 2010), thus adding up to 6 more measurements per star\footnote{Data compiled from: 
Jenkins et al. (2011); Weise et al. (2010); Houdebine (2010, 2008); Schr\"{o}der et al. 
(2009); Mishenina et al. (2008); Scholz et al. (2007); Desidera et al. (2006); Valenti \& 
Fischer (2005); Santos et al. (2004); Nordstrom et al. (2004).}. The G\l\c{e}bocki \& 
Gnaci\'{n}ski (2005b) method of merging calibrations was not reproduced, but the weights 
$w$ they attribute to different methods of line fitting were adopted. The weighted 
standard error on the mean is then $\sigma/\sqrt{N_{eff}}$, for an effective number of 
observations $N_{eff} = (\Sigma(w))^2/\Sigma(w^2)$. For values differing from the mean by 
$\delta$, $\sigma = \sqrt(\Sigma(w\delta)/\Sigma(w))$. The number of velocities included 
is 5 to 14, with $N_{eff}$ of 4.3--13.5, except for the very slow rotator 
GJ~581, whose $v \ sin \ i_*$ (Marcy \& Chen 1992) does not constrain the stellar 
inclination. Overall, some differences in $v \ sin \ i_*$ between different catalogues 
were confirmed; omitting particular datasets shifts the means by up to $\approx1.5\times$  
the standard error.

Periods $P$ are found from tracking variability associated with surface inhomogeneities, 
such as the data obtained under the long-running Mount Wilson Project. Such results are 
sparse, and limit our analysis to 9 nearby late-type (F9--M3) stars. Uncertainties and 
intrinsic variations in $P$ are generally recorded as small, at $\sim$5~\%. Hartman et 
al. (2011) investigated reliability of period extraction in a star survey including BY 
Dra rotational variables (including HD~166, HD~30495, $\epsilon$~Eri, GJ~581 and AU Mic 
here), and only the latter two M-stars have amplitudes in the 0.01-0.02~mag range that is 
of concern. Of these, only AU Mic rotates fast enough for useful analysis here, and the 
period was derived from a set of 10 light curves (Messina et al. 2001). More ambiguous 
periods could however arise in cases of differential surface rotation and/or temporal 
changes. The most extreme case noted here is HD~30495, where Baliunas et al. (1983) found 
a period of 7.6 days, in contrast to 10.5-11.5 days in more recent data (Gaidos et al. 
2000). To illustrate this `worst case' uncertainty, using the low period value and the 
lower bound in $v \ sin \ i_*$ would give a stellar inclination at the -2.3$\sigma$ bound 
compared to the Table~1 solution.

\hspace*{-5cm}
\begin{figure} 
\label{fig2} 
\includegraphics[width=82mm,angle=0]{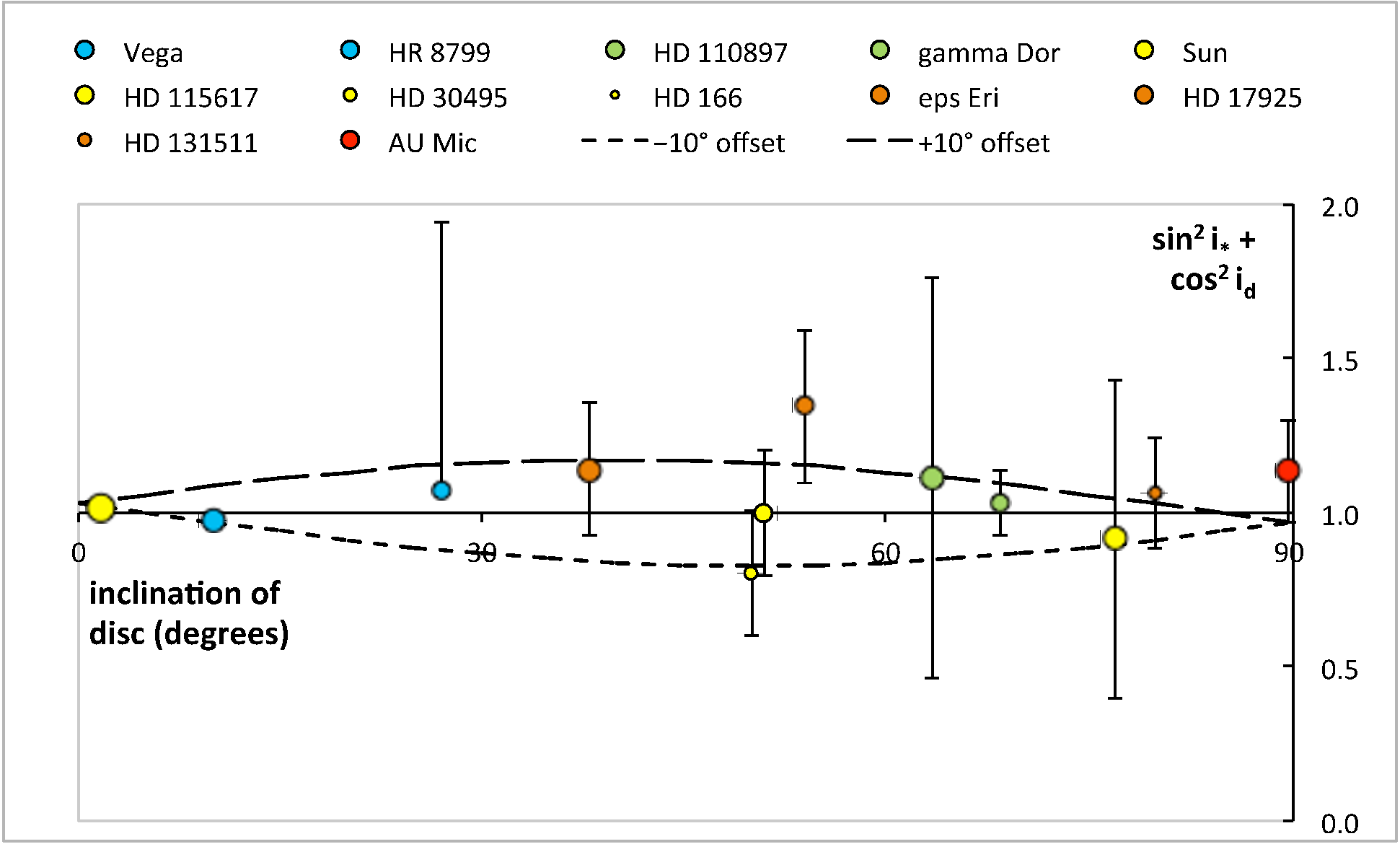} 
\caption{Plot of $sin^2 i_*+cos^2 i_d$ against disc inclination, where y-values $\neq 1$ indicate 
that the disc and star are misaligned. Symbols are ordered as in Table~1, with colours according 
to type (from blue for A to red for M) and smaller sizes for greater distances. X-axis error 
bars are omitted for clarity. The curves illustrate how $sin^2 i_*+cos^2 i_d$ varies with disc 
inclination, when the star has a {\it relative} tilt of $\pm10^{\circ}$.  
}
\end{figure}

Periods can be estimated from relations linking main-sequence spin-down to decline in 
chromospheric activity (e.g. Mamajek \& Hillenbrand 2008). However, for more 
FGK discs resolved in DEBRIS/DUNES, this method showed a problem of $sin \ i_* > 1$ for 
30~\% of stars; Kennedy et al. (2013) discuss how $i_*$ can be robust if it is small. 
An advantage of observed periods is that spin-down for Solar-type stars is rather 
well understood (Barnes 2007), and so `gyrochronology' ages have be found (Vican et al. 2012); 
this confirmed the unbiased nature of our survey targets.


\section{Results}

Results are listed in Table~1. Star-disc inclination differences $\Delta i$ are typically 
close to zero, albeit with large errors where $v \ sin \ i_*$ is low. In the seven 
best-defined cases, the star-disc systems appear co-planar within 5$^{\circ}$ on average, 
with only the Vega system potentially misaligned (by $5.5 \pm 2.5^{\circ}$). This small tilt 
would be similar to the Sun's inclination versus the Kuiper belt, which for an external 
observer in an ecliptic co-ordinate frame would be $\Delta i = 5.6^{\circ}$. A 
potentially misaligned system is the planet-host HR~8799, which has only a lower limit to 
$i_*$ from asteroseismology and $\Delta i \ga 3^{\circ}$; if the star is 
far from pole-on it will not be co-planar with the disc.

The survey outcome is similar to the null result of Watson et al. (2011), from eight 
stars. The joint sample now covers 16 stars with useful $\Delta i$ values, with 
three-quarters of these now observed uniformly by Herschel. Given the null results, no 
stellar property (Table~1) is noteworthy -- unlike the situation for close-in planets, 
where e.g. a link with the proportion of the star that is convective has been suggested 
(Winn et al. 2010). For completeness, we note that a binary-star system is known with a 
highly misaligned (circumpolar) debris disc (Kennedy et al. 2012), but here our stars 
are single, except for the spectroscopic binary HD~131511.

To assess any {\it mean} tilt present, we use the measure $sin^2 i_* + cos^2 i_d$, which 
diverges from unity if the disc and star are misaligned. This is more statistically 
convenient than $\Delta i$, as measurement errors in $sin \ i_*$ and $i_d$ can be assumed 
to be normally distributed. The errors can be written as $\delta(sin^2 \ i_*) = 2 \ sin \ 
i_* \ \delta(sin \ i_*)$ and $\delta(cos^2 \ i_d) = 2 \ cos \ i_d \ sin \ i_d \ 
\delta(i_d)$ and combined quadratically. The mean value of $sin^2 i_* + cos^2 i_d$ 
(excluding the Sun) is then 1.06 with a standard error of $\pm 0.04$, consistent with no 
misalignment at the 1.5$\sigma$ level. Figure~2 illustrates this by plotting $sin^2 i_* + 
cos^2 i_d$ for the whole sample. The value obtained for a particular star-disc tilt 
depends on viewing angle, and the over-plotted curves illustrate example relative tilts. 
These curves at $\pm10^{\circ}$ encompass all plotted stars within their errors, 
suggesting the mean tilt is within this range.

\section{Discussion}

The generally good alignment of stars with their debris discs is in marked contrast 
to the situation for close-in planets. The cases can not be absolutely compared, 
because transit data yield inclination differences up to 180$^{\circ}$, versus a 
0-90$^{\circ}$ range for disc-star alignment, while neither method is fully 3-D 
(lacking the orientation of the stellar pole). However, approximately a third of 
Rossiter-McLaughlin detections have $\Delta i$ of 30-150$^{\circ}$, for example 
(Brown et al. 2012), while here there are no good candidates for this magnitude of 
misalignment. This suggests that dynamical effects near the star do not operate on 
the outer system planetesimals. 

A few debris-host stars also have imaged planet-candidates, at semi-major axes of 
15-180~AU. These systems suggest planet-disc co-planarity, as well as the star-disc 
alignments. HR~8799~b has an orbital plane inclined at 13--23$^{\circ}$ (Lafreni\`{e}re 
et al. 2009) versus our 17--37$^{\circ}$ for the disc plane; Fomalhaut~b's orbit is 
estimated at $17\pm12^{\circ}$ from the ring plane (Kalas et al. 2013); $\beta$~Pic~b 
(Lagrange et al. 2012) is thought to have perturbed the inner-disc plane to align close 
to its orbit; and $\epsilon$~Eri~b (unconfirmed, at $\sim$3~AU) has a nominal astrometric 
orbit within $\sim 10^{\circ}$ of the outer debris belt plane (Greaves et al., in prep.). 
This suggests different forces at work than on close-in planets, or binary stars, where 
orbits and spin axes tend to misalign at separations $\ga30-40$~AU (Hale 1994). The 
`regime of coplanarity' is hard to define, though Figueira et al. (2012) have suggested 
that HARPS plus Kepler detection statistics may point to co-planarity of multiple planets out 
to about 0.3~AU.

To make further progress, it would help to discover transiting-planet-plus-disc systems 
(Hebb et al. 2007), as well as to resolve tilts within more multiple-belt systems like 
$\beta$~Pic. Generally, models where external encounters affect the alignment of outer 
components of the system (Bate et al. 2011; Thies et al. 2011) seem unlikely, as 
planets and discs at different radii should be differently perturbed, while here 
we find examples of stars aligned with both disc and planets over tens-of-AU scales. We 
note especially the case with the most 3-D information, the Fomalhaut system, where the 
orientation of the stellar pole is orthogonal to the disc plane (Le Bouquin et al. 2009), 
and Fomalhaut~b's orbit is close to the plane of the debris ring (Kalas et al. 2013). 

\section*{Acknowledgments} 

{\it Herschel} is an ESA space observatory with science instruments provided by 
European-led Princpal Investigator consortia and with important participation from NASA. 
This work was supported by ERC grant 279973 (GMK, MCW) and Spanish grant AYA 2011-26202 
(CE, JPM, JM).

\bsp

\label{lastpage}

\end{document}